\documentstyle[12pt,epsf]{article}
\newcommand{\D}{\displaystyle}
\newcommand{\T}{\textstyle}

\newcommand{\SSt}{\scriptscriptstyle}
\begin{document}
\begin{center}
{\large\sc  Stochastic Schemes of Dielectric Relaxation\\
in Correlated-Cluster Systems}
\\[30pt]
{\bf Andrew K.~Jonscher}\\[10pt]
Royal Holloway, University of London\\
Egham, Surrey, TW20 0EX, UK\\
e-mail: jonscher@lynwood.u-net.com\\[10pt]
{\bf Agnieszka Jurlewicz}\\[10pt]
Hugo Steinhaus Center for Stochastic Methods and 
Institute of Mathematics,\\ Wroc{\l}aw University of Technology\\
Wyb.~Wyspia{\'n}skiego 27, 50--370 Wroc{\l}aw, Poland\\
e-mail: A.Jurlewicz@im.pwr.wroc.pl\\[10pt]
{\bf Karina Weron}\\[10pt]
Institute of Physics, Wroc{\l}aw University of Technology\\
Wyb.~Wyspia{\'n}skiego 27, 50--370 Wroc{\l}aw, Poland\\
e-mail: karina@rainbow.if.pwr.wroc.pl\\[20pt]
\end{center}
\begin{abstract}
Unlike the classical exponential relaxation law, the widely prevailing
universal law with its fractional power-law dependence of
susceptibility on frequency cannot be explained in the framework of any
intuitively simple physical concept. The resulting constancy of the
ratio of the imaginary to the real parts of the complex susceptibility,
known as the ``energy criterion'', has a pleasing simplicity but the
understanding of its origins needs a special theoretical treatment. A
fresh light on the stochastic nature of the dielectric relaxation 
has been shed by a novel stochastic approach
introduced in the last decade. Since the theoretical analysis involved
is rather unfamiliar, the aim of this paper is to give some useful
comments and suggestions which should help to follow in details the
proposed stochastic scheme of relaxation leading to the well-known
empirical responses. We justify the universality of the power-law
macroscopic response as well as Jonscher's screening and energy
criterion ideas, and we give a new basis to the research into the
significance of relaxation processes. \\[1ex]
\end{abstract}
\newpage
\section{Introduction.}

Dielectric relaxation in solids, a process commonly defined as an
approach to equilibrium of a dipolar system driven out of equilibrium by 
a step or alternating external electric field, represents one of the
most intensively researched topics in experimental and theoretical
physics, see e.g.~[1-21].
Wide-ranging experimental information resulting from the studies of the
basic dielectric characteristics (i.e., the time decay of the depolarization 
current $i(t)$ and the frequency-dependent complex dielectric permittivity 
$\epsilon (\omega)$ or susceptibility $\chi (\omega)\propto
\epsilon(\omega)-\epsilon_{\infty}$, where $\epsilon_{\infty}$ is the
asymptotic value of the dielectric permittivity $\epsilon(\omega)$ at
high frequencies) 
has led to the conclusion that the classical phenomenology of relaxation 
breaks down in complex materials.
It has been found that the Debye behaviour, represented by the
exponentially decaying relaxation function 
\begin{equation}
\label{D}
\phi (t)=\exp(-\omega_{p} t)
\end{equation}
yielding
$$
i(t)\propto f(t)= -\D{d\phi (t)\over dt}=
\omega_{p}\exp\left (- \omega_{p} t\right ) 
$$
and
\begin{equation}
\label{D1}
\chi(\omega)\propto\phi^{*}(\omega)= 
\int\limits_{0}^{\infty} e^{-i\omega t} f(t)dt =
\D{1\over 1+i\omega / \omega_{p}}
\end{equation}
(where the constant $\omega_{p}$ denotes the loss peak frequency)
is hardly ever found in nature and that the deviations from it for
many dielectrics may be relatively large 
\cite{jonscher83, havriliak, jonscher96}.
A class of systems exhibiting
the non-Debye relaxation patterns includes various complex materials
such as supercooled liquids, amorphous semiconductors and insulators,
polymers, disordered crystals, molecular solid solutions, glasses, etc.

For a long time effort was being diverted to a
purely qualitative representation of the shape of the non-Debye dielectric
functions $\phi (t)$ or $\phi^{*}(\omega)$ 
in terms of certain mathematical expressions without in any
way going into the physical significance of these representations. 
It turns out that all dielectric data are characterised well enough 
by a few empirical functions
\cite{bottcher, jonscher83, havriliak, jonscher96}.
The time-domain relaxation data usually are 
fitted by means of the stretched exponential relaxation function 
\begin{equation}
\label{KWW}
\phi (t)=\exp \left (-(\omega_{p} t)^{\alpha}\right )
\;\;\;{\rm with}\;\;0\!<\!\alpha\!<\!1
\end{equation}
known as the Kohlrausch-Williams-Watts (KWW) function and coincident
with the Debye case (\ref{D}) for $\alpha=1$.
The most popular analytical expression applied to the complex susceptibility
data is given by the Havriliak-Negami (HN) function 
\begin{equation}
\label{HN}
\phi^{*}(\omega)=
\D{ 1 \over \left ( 1+(i\omega/\omega_{p})^{\alpha}
\right )^{\gamma}}\;
\end{equation}
where $0\!<\!\alpha, \gamma\!<\! 1$. For $\alpha\!=\!1$ and
$\gamma\!<\!1$, formula (\ref{HN}) takes the form known as 
the Cole-Davidson (CD) 
function; for $\gamma\!=\!1$ and $\alpha\!<\!1$ it 
takes the form of the Cole-Cole (CC) function, and for $\alpha\!=\!1$ and 
$\gamma\!=\!1$ one obtains the classical Debye form (\ref{D1}).

We repeat that these various model functions are essentially attempts
to characterise the observed behaviour 
without in any way indicating the physical mechanisms
involved. To that extent, we note that the problem of 
understanding the nature of dielectric relaxation is as yet largely
open since we do not have a sufficiently general method of approach to
the relaxation processes. At the same time we note certain fundamental
features of relaxation response which make it clear that we are faced
with some very general processes into which we require a much deeper
insight.

The HN, CC, CD, and KWW model functions all show the high-frequency
limit of the response characterised by the fractional power law 
\cite{bottcher, jonscher96}:
\begin{equation}
\label{fpl}
\chi '(\omega)=\tan \left (\D{n\pi \over 2}\right )\; 
\chi '' (\omega)\propto \omega^{n-1}
\;\;\;\mbox{\rm for some}\;\;0<n<1
\end{equation}
which has the straightforward
consequence of the constant ratio of the imaginary to the real components 
of the dielectric susceptibility $\chi (\omega)=\chi'(\omega)-
i\chi''(\omega)$:
\begin{equation}
\label{encr}
\D{\chi ''(\omega) \over \chi ' (\omega)} =
{\rm cot}\left (n\D{\pi\over 2}\right )
\;\;\;{\rm for}\;\omega\gg\omega _{p}.
\end{equation}
Let us note that this relation implies that the ratio of energy lost per 
radian to the energy
stored at the peak is independent of frequency. This very simple and
intuitively evident property is characteristic of the prevailing 
universal response of most dielectric materials. We call this the
``energy criterion'' and it is neither known nor predicted by any of
the accepted theories of relaxation.

Important for our discussion of relaxation phenomena is the
concept of residual loss \cite{jonscher2001}
which remains after the removal of such strongly frequency-dependent
processes as the direct-current contribution going as $\omega^{-1}$, or
low-frequency dispersion where loss is proportional to $\omega^{n-1}$
with $n\to 0$, or dipolar loss which shows pronounced peaks. The
residue left is broadly frequency-independent or only weakly dependent
over extended ranges of frequency which may be approximated by the
limit $n\to 1$. Such a limit corresponds to frequency-independent 
components of the dielectric susceptibility, $\chi'(\omega)$ and 
$\chi''(\omega)$, with their ratio tending to infinity  
which is
impossible practically but can be seen experimentally in various
approximations. Figure 1 shows a compilation of data for different
materials taken from much wider sets for different temperatures. These
data are representative of a very wide range available in the published
literature and are quoted here purely by way of examples. The data shown
correspond to low or relatively low losses so that the real part of
the permittivity $\epsilon'(\omega)$ does not vary by a large amount
over the frequency range in question; and it is therefore possible to
normalise the data by dividing $\chi''(\omega)$ by the corresponding
$\epsilon'(\omega)$, thus defining the loss tangent
$\tan\delta=\chi''(\omega)/\epsilon'(\omega)$, which gives absolute
values of loss for the various samples while retaining the frequency
dependence of $\chi''(\omega)$.
The essential conclusion, coming from the analysis of the data presented
in Figure~1, is that the residual loss follows in some
cases law (\ref{fpl}) while in other cases it is essentially 
``flat'' in frequency with minor perturbations. It is interesting to
note that the addition of 17\% of carbon black to
polyethylene causes a rise of $\tan\delta$ by one order of magnitude
but hardly any change of the frequency dependence. It is also
noteworthy that the total range of $\tan\delta$	covers at most three
decades for materials for which the direct-current conductivity varies
by many orders of magnitude. It is not profitable to fit these data to
any particular law since they are the result of the interplay of minor
accidental factors superimposed on the prevailing broad ``flatness''.
A theoretical justification for the existence of the fractional power
law of type (\ref{fpl})	with the exponent $n$ arbitrarily close to unity 
would
be sufficient for the understanding of all flat losses, any minor
deviations being purely accidental.

In our theoretical approach to the analysis of the universal-response 
characteristics (\ref{fpl}) we make use of
the concept of dipolar screening introduced by Jonscher
\cite{jonscher97}.
In the framework of this concept, the effectiveness of
screening depends on the relative magnitudes of the 
thermally-activated-dipole density $N_{d}$ with an energy $W$, 
$N_{d}\propto \exp (-W/kT)$,
and the ``critical'' density $N_{1}\propto kT/ \mu^{2}$
where $\mu$ is the dipole moment of the dipole being
screened. 
The theory of dipolar screening
predicts that for low dipolar density, $N_{d}\ll N_{1}$, the number of
dipoles within the field of any one dipole increases almost
exponentially, so that their behaviour is strongly collective and
the etire system behaves in the universal manner (\ref{fpl}). 
On the other hand, for a
high dipolar density, $N_{d}\gg N_{1}$, screening is effective, meaning
that dipoles do not ``see'' their neighbours and behave individually in
a Debye-like manner. This is summarized in Table 1.
The situation is in many respects similar to the classical screening by
charged particles but the onset of screening is much more rapid with a
fall in the dipolar density $N_{d}$.

A considerable effort that has been devoted in the past to finding a
theoretical explanation of the empirically observed results points on 
the two most widespread and at the same time least understood properties of
the relaxation responses:
\begin{itemize}
\item the existence of the characteristic property (\ref{fpl})
with its limiting form of virtually frequency-independent loss;
\item the fact that this property of the relaxation responses 
is common to a very
wide range of materials with very different physical and chemical
interactions.
\end{itemize}
As a consequence, in theoretical attempts to model relaxation it has
been commonly assumed that the empirical relaxation laws reflect a kind
of general behaviour which is independent of the details of examined
systems.  This idea has stimulated the proposal of several relaxation 
mechanisms that differ mainly in the interpretation of the relaxation 
function. In recent attempts to find the origins of the non-Debye
relaxation patterns the idea of
complex systems as  the  ``structures with variations'' \cite{gold}
that are characterised through a large diversity of elementary units
and strong interactions between them is of special importance.  The
evolution in course of time of physical properties of a complex system 
is nonpredictable or anomalous \cite{guerra}, and the main feature of
all the dynamical processes in such a system is their stochastic
background. In the framework of statistical models (see
e.g.~[4-6, 8, 9, 11-14, 17-21, 26-31]
the fact that 
the large scale behaviour of complex systems shows
universality, i.e., that it is to some extent independent of the precise
local nature of the considered system, should come as no surprise.
Intuitively, one expects ``averaging principles'' like the law of large
numbers to be in force. However, it turns out to be very hard to make
this intuition precise in concrete examples of stochastic systems with 
a large number of locally interacting components.  

The empirical facts stress the need for a completely novel approach to the
modelling of the dielectric relaxation and, to a certain
extent, also mechanical relaxation, photoconduction, photoluminescence 
and chemical reaction kinetics (sharing some common features, see 
\cite{jonscher96}).
 The need to understand the connections between the macroscopic
property (\ref{fpl}) of the relaxing complex system and the statistical 
properties of individual molecular or dipolar species requires the
introduction into the relaxation theory of advanced methods of
stochastic analysis. 
As shown  by us [8, 32-34],
a general formalism of limit 
theorems of probability theory plays an important role 
in constructing tools to
relate the local random characteristics of the complex system to the
empirical, deterministic relaxation laws, regardless of the specific
nature of the system considered. The significance of the present paper
lies in the fact that no one has clarified in a simple and plausible
way, let alone one based on a stochastic argument, why  the
universal relation should exist at all and why the residual loss is
such slowly variable function of frequency. Our approach provides a
rigorous explanation for the most widely observed form of frequency
dependence of the permittivity (or susceptibility) and thereby opens up
a new and very powerful way of interpreting relaxation phenomena not
only in the dielectric context.

The main objective of this paper is to
focus on the approach to relaxation in the framework proposed in
\cite{kwaj2001,ajkw2001} as consistent with Jonscher's screening and
energy criterion ideas and providing their strict mathematical
formulation. We also 
bring to light the spatio-temporal scaling conditions 
hidden behind all the well-known empirical responses. 
For the reader's convenience, the mathematical details
neccessary for the stochastic construction of the effective representation
of a relaxing complex system are followed by extended comments.

In Section 2 we introduce the basic mathematical concept underlying the
stochastic transition of a complex system from its initially imposed
state. We show that the relaxational properties of the entire system
can be represented  by means of a random effective relaxation rate
which contains information on the internal stochastic structure of the
investigated system. In Section~3 we point to the origins of the
statistical properties of the effective relaxation rate. We discuss the
role of limit theorems of probability theory as the mathematical
technique which allows us to derive the explicit relaxation formulas
even with rather restricted information on local random properties of
the system. In Section~4 we relate the statistical
conditions yielding the well-known empirical responses to the 
spatio-temporal  scaling properties of the relaxing complex system, and
we show that they underlie 
Jonscher's energy-criterion hypothesis. The last section contains
conclusions.


\section{Transition and survival probabilities}
{\bf (i)} 
Let us consider a complex physical system containing identical objects 
undergoing irreversible transitions from state $A$, imposed at time $t=0$,   
to state $B$ at random instants of time.
States $A$ and $B$ differ in some physical parameter, so that the
transition $A\rightarrow B$ is defined as the change of this particular
parameter (changes in all other parameters may also have an influence
on the transition). Let us choose one of the objects. Consider the 
conditional probability $p(t, dt)$ that this object will 
undergo the transition during the time interval $(t, t+dt)$ if the transition
has not occured before time $t$, i.e.
\begin{equation}
\label{7}
p(t, dt)=\Pr (t\leq \theta\leq t+dt|\theta\geq t)
\end{equation}
where $\theta$ is the random waiting time for the transition of the
chosen object. The 
conditional probability $p(t, dt)$ defined in (\ref{7}) can be
expressed in a form
$$
p(t, dt)=-\D{\Pr (\theta\geq t+dt)-\Pr (\theta\geq t)
\over \Pr (\theta\geq t)}
$$
where $\Pr (\theta\geq t+dt)$ and $\Pr (\theta\geq t)$ are the
survival probabilities, i.e. the probabilities that the considered
object will remain in state $A$ until time $t+dt$ and $t$,
respectively. 
The survival probability of the object can be expressed as 
$$
\Pr (\theta\geq t) = 1- p(t)
$$
where
$$
p(t) = \Pr (\theta < t)
$$
is the waiting time distribution of the object, i.e. the total
probability of its transition $A\to B$ until time $t$.
One can rewrite the survival probability in the form 
\begin{equation}
\label{8}
\Pr (\theta\geq t)=
\exp (\T -\int\limits_{0}^{t} r(s)ds)
\end{equation}
which is dependent on a non--negative quantity $r(s)$ called the
intensity of transition \cite{vankampen}. This quantity is
time--dependent, in general; and moreover, 
because of random impacts affecting each object, for different 
objects in the system it can take various values at the same instant of 
time \cite{berlin93}.

The survival probability $\Pr (\theta\geq t)$ can be derived if 
one knows the explicit form of the intensity $r(s)$. 
For the time independent intensity $r(s)=b_{0}=const$ one gets
$$
\Pr (\theta\geq t)=\exp (\T -b_{0}t)
$$
what recovers the classical exponential evolutionary law for the 
object; the value $b_{0}$ determines the relaxation rate of the
transition process. 
If, on the contrary, the intensity of transition is essentially
time--dependent, then 
the evolutionary law for the object is of the nonexponential form. In this 
case the resulting relaxation rate is not
directly given by the intensity of transition as it is in the exponential
case. In fact, taking into account the peculiarity of local random
environment of an object,  one should assume that, in general, the survival
probability $\Pr (\theta\geq t)$ has the form of the weighted average of 
an exponential decay with respect to the probability distribution 
$F_{\beta}(b)$ of the relaxation rate of the object
\begin{equation}
\label{9}
\Pr (\theta\geq t) = 
\int\limits_{0}^{\infty} e^{\T -bt}dF_{\beta}(b).
\end{equation}
In other words, the
relaxation rate of the object is the random variable $\beta$ such that 
the total survival probability has the form 
\begin{equation}
\label{10}
\Pr (\theta\geq t) = \left \langle \exp (-\beta t)\right \rangle 
\end{equation}
where the mean value  $\langle\cdot\rangle$ is taken 
with respect to the relaxation-rate probability distribution
$F_{\beta}(b)$.
Expressions (\ref{9}) and (\ref{10}) are 
a generalisation of the Debye relaxation 
for which the
relaxation rate $\beta$ takes a constant value, say $b_{0}$, with
probability 1, $\Pr (\beta=b_{0})=1$. Then $\beta$ has a degenerate
probability distribution, $dF_{\beta}(b)=\delta(b-b_{0})db$ (where
$\delta (\cdot)$ denotes the Dirac delta function), and the
integral on the right--hand side  of (\ref{9}) equals
$$
\int\limits_{0}^{\infty} e^{\T -bt}\delta(b-b_{0})db=\exp (-b_{0}t).
$$ 
In general, in a complex system 
the probability distribution $F_{\beta}(b)$ and,
consequently, the survival probability $\Pr (\theta\geq t)$ of an
object are of unknown forms.

From equation (\ref{8}) the intensity of transition from an
initial state, $r(t)$, is related to the total survival probability 
$\Pr (\theta\geq t)$ of the object as follows
\begin{equation}
\label{11}
r(t)= -\D{d\over dt}\ln \Pr (\theta\geq t).
\end{equation}
Comparing (\ref{11}) and (\ref{9}) we obtain the relationship between
the time--dependent intensity of transition and the relaxation rate
distribution
$$
r(t) = -\D{d\over dt}\ln \int\limits_{0}^{\infty} 
e^{\T -bt}dF_{\beta}(b).
$$
The above mentioned lack of information about the 
relaxation rate distribution $F_{\beta}(b)$ yields that the corresponding 
intensity $r(t)$ is of unknown form. 
The explicit form of $r(t)$ depends on the characteristics of 
the random environment around the examined object and the rules 
needed to specify the sets of deterministic and stochastic
parameters.
\newpage
\noindent
{\bf (ii)}
Consider (as before) a system of $N$ objects, each waiting for transition 
$A\rightarrow B$ for some  random time. For the $i$th object, $1\leq i\leq
N$, let the
waiting time be denoted by $\theta_{\SSt iN}$. 
The notation here (i.e.~index ``$iN$'') emphasizes the impact of the 
system size $N$ on the behaviour of each individual object that 
has to be taken into account. 
The non--negative waiting times $\theta_{\SSt 1N},\ldots ,
\theta_{\SSt NN}$ form a sequence of independent, identically distributed 
random variables.  The waiting time distribution 
$p_{iN}(t)=\Pr (\theta_{\SSt iN} < t)$ is 
hence the same for each~$i$
and is equal to some function $F_{\theta} (t)$ that may depend on $N$. 

The objects undergo transition in a certain order that can be reflected 
in the notion of order statistics \cite{feller}
$\theta_{(1)}\leq\ldots\leq\theta_{({\SSt N})}$, 
which is simply a nondecreasing rearangement of times
$\theta_{\SSt 1N},\ldots ,\theta_{\SSt NN}$. Traditionally,
$\theta_{(l)}$ is called the $l$--th order statistics of sample 
$\theta_{\SSt 1N},\ldots ,\theta_{\SSt NN}$. Note that 
$\theta_{(1)}=\min(\theta_{\SSt 1N},\ldots ,\theta_{\SSt NN})$ and 
$\theta_{({\SSt N})}=\max(\theta_{\SSt 1N},\ldots ,\theta_{\SSt NN})$.

For a fixed size $N$ of the system, the ratio of objects not transformed
up to time $t$ is equal to
\begin{equation}
\label{12}
1 - \D{\eta_{\SSt N}(t)\over N}
\end{equation}
where $\eta_{\SSt N}(t)$ denotes an unknown (random!) number of objects 
already transformed at time $t > 0$. For $l=0, 1,\ldots N$
the events $\{\eta_{\SSt N}(t) = l\}$ that up to time $t$ exactly 
$l$ transitions occured in the system 
can be expressed via order statistics in the following way:
\begin{equation}
\label{13}
\begin{array}{lll}
\{\eta_{\SSt N}(t) = 0\}& =& \{\theta_{(1)}>t\},\\
\{\eta_{\SSt N}(t) = l\}& =& \{\theta_{(l)}\leq
t,\;\theta_{(l+1)}>t\}\;\;\;
{\rm for}\;\; l=1,\ldots N\!-\!1,\\
\{\eta_{\SSt N}(t) = N\}& =& \{\theta_{({\SSt N})}\leq t\},
\end{array}
\end{equation}
and the probability of such events is given by the Bernoulli model
\begin{equation}
\label{14}
\Pr (\eta_{\SSt N}(t) = l) = {N\choose l}
\left [p_{\SSt iN}(t)\right ]^{\T l}\left [1-p_{\SSt iN}(t)\right ]^{\T
{N\!-\!l}},\;l=0, 1,\ldots N.
\end{equation}
Thus the random number $\eta_{\SSt N}(t)$ has the Bernoulli (binomial)
distribution ${\cal B}(N, p)$ with parameter $p = p_{\SSt iN}(t)$. 
Consequently, the fraction in (\ref{12})
is a random variable. However, it follows from the strong law of large 
numbers \cite{viniotis} that for $N$ large enough 
\begin{equation}
\label{15}
\D{\eta_{\SSt N}(t)\over N} \approx \left\langle
\D{\eta_{\SSt N}(t)\over N}\right\rangle = p_{\SSt iN} (t) = F_{\theta}(t)
\end{equation}
i.e. the random nature of the fraction in (\ref{12}) vanishes for large
$N$. Hence (\ref{12}) is asymptotically equal to 
$(1-p_{\SSt iN}(t))$ which is the survival probability 
$\Pr (\theta_{\SSt iN}\geq t)$ of any single object:
\begin{equation}
\label{16}
1 - \D{\eta_{\SSt N}(t)\over N} \approx 1-p_{\SSt iN}(t)
=\Pr (\theta{\SSt iN}\geq t).
\end{equation} 

The basic property
of the relaxation function $\phi (t)$ is its monotonic decrease
from 1 at $t=0$ to 0 as $t\to\infty$. This is in fact a property of the 
survival probability and therefore (\ref{12}) has been used
as a definition of the relaxation function in several models. 
Yet (\ref{12}) can be used to describe the evolution 
of the entire system only if the behaviour of the system is represented by any 
individual object (from those forming the system). Unfortunately, this
strict condition seems to contradict the idea of complexity of the 
investigated systems, and hence models defining the survival
probability of a complex system as in (\ref{16})
do not capture the nature of relaxation phenomena.
Moreover, the relation (\ref{15}) holds for any arbitrarily chosen
form of the waiting time distribution (e.g. 
$F_{\theta}(t)=0$ for $t<0$, $=t$ for $0\leq t \leq 1$, and $=1$ for
$t>1$) so that the ratio in (\ref{15}) is not uniquely 
determined. Therefore the attempts leaving the probabilistic 
analysis of the
irreversible stochastic transitions in complex systems at this stage
can only propose the form of distribution $F_{\theta}(t)$ fitting the
data most exactly, see e.g.~\cite{berlin93,berlin94}. 
They do not explain the observed fractional power law 
(\ref{fpl}) indicating strictly limiting
properties of the survival probability of the initial state of a complex
system.\\[8mm]
{\bf (iii)}
The considerations of irreversible stochastic transitions in complex systems
show that the behaviour of the system as a whole, in general, cannot be
attributed to any chosen object forming the system 
\cite{berlin93,kwaj93,berlin94,kwkotulski}. Crucially
relevant to this statement is the question of a proper mathematical
construction of an ``averaged'' imaginary object respresenting the entire 
system. As it follows from (\ref{8}), the intensity of transition from
an initial state for any (real or imaginary) object depends on 
its survival probability. Let us 
denote the  survival probability of an imaginary object representing 
the system as a whole by $\Pr (\tilde\theta_{N} \geq t)$ and by
$\tilde\theta_{N}$ 
the effective waiting time for the entire system. 
Since the considered imaginary object represents the whole system, 
the survival probability $\Pr (\tilde\theta_{N} \geq t)$ is the probability 
that the transition of the system as a whole from its initial
state (imposed by external constraints at $t=0$) has not happened prior to a
time instant~$t$. This mathematical quantity is defined as the probability
that there is no transition occuring in the system up to time $t$ so
that
$$
\Pr (\tilde\theta_{N} \geq t) = \Pr (\eta_{N}(t)=0).
$$
By means of the Bernoulli scheme (\ref{14}) with $l=0$, 
the survival probability of the system 
is the product of $N$ factors, each asymptotically equal to (\ref{12}), 
see equation (\ref{16}):
$$
\Pr (\tilde\theta_{N} \geq t) = \Pr (\eta_{N}(t)=0) = (1-p_{iN}(t))^{N}.
$$
On the other hand, by means of the order statistics 
$\{\eta_{N}(t)=0\} = \{\theta_{(1)}\geq t\}$, see equation (\ref{13}); 
and the survival probability of the entire system 
$$
\Pr (\tilde\theta_{N} \geq t) = \Pr (\eta_{N}(t)=0) = \Pr(\theta_{(1)}\geq t) = 
\Pr (\min (\theta_{1N},\ldots ,\theta_{NN})\geq t)
$$
is just the probability that the first passage
of the system from the initial state has not happened before time~$t$. 

Since the waiting times $\theta_{1N},\ldots ,\theta_{NN}$
are independent and identically distributed, we have
$$
\Pr (\tilde\theta_{N} \geq t) = \Pr(\theta_{1N}\geq t)\ldots\Pr(\theta_{NN}\geq 
t). 
$$
Assuming (\ref{10}) for each $\theta_{iN}$, i.e.
$$
\Pr (\theta_{iN}\geq t) = \left \langle \exp (-\beta_{iN} t)\right
\rangle 
$$
where random variables $\beta_{1N},\ldots,\beta_{NN}$ are also independent
and identically distributed,
we obtain that the survival probability of the system is of the form 
analogous to (\ref{10}):
\begin{equation}
\label{17}
\Pr (\tilde\theta_{N} \geq t) =
\left\langle \exp(-t\tilde\beta_{N})\right\rangle
\end{equation}
where 
\begin{equation}
\label{18}
\tilde\beta_{N} = 
\sum\limits_{i=1}^{N}\beta_{iN}.
\end{equation}
The random variable $\tilde\beta_{N}$ can be considered as the
effective relaxation rate representing the entire 
system; and, in accordance with the rate-theory concept (see
e.g.~\cite{vlad}), individual relaxation rates $\beta_{iN}$'s 
are hence some contributions to the total
rate $\tilde\beta_{N}$.

In fact, in any complex dielectric system under a weak external electric field 
only a part of the total number $N$ of dipoles, referred to as ``active''
dipoles, is directly governed by changes of the field.
The exact number of such dipoles depends on temperature, density of the 
dipolar species in the system, and interactions between them, and
it is usually unknown. (In some cases it may take the value of the total 
number of the dipoles in the system.) 
Therefore, there can be only a part of dipoles contributing to the
effective relaxation rate, and it is reasonable to modify formula (\ref{18})
taking some unknown number $\nu_{N}$ instead of exactly $N$ components:
\begin{equation}
\label{19}
\tilde\beta_{N} =  
\sum\limits_{i=1}^{\nu_{N}}\beta_{iN}.
\end{equation}

\section{Origins of the random effective relaxation rate}

The relaxation function $\phi (t)$, which is the probability that the
initially imposed state of a macroscopic system survives by time $t$, 
is well known to be expressed as the weighted average of an exponential
decay with respect to the distribution of the effective relaxation rate
$\tilde\beta_{N}$ \cite{vlad,klafter,fuente},
and its explicit form depends on statistical properties of the rate. 
In view of (\ref{17}),
we can use the survival probability $\Pr (\tilde\theta_{N}\geq t)$ as a
definition of the relaxation function and we obtain in this case that 
\begin{equation}
\label{20}
\phi (t)=\left\langle \exp(-{\tilde \beta}_{N} t) \right\rangle
\end{equation}
where $\tilde\beta_{N}$ is given by (\ref{19}). 

The evaluation of $\nu_{N}$, the number of components in (\ref{19}), 
which is in
agreement with the nature of the relaxation phenomenon, has been
presented in \cite{kwaj2001, ajkw2001}. It is based on the following
idea:

Depending on the strength of screening, the active dipoles in the
complex system may 
``see'' each other to some extent. The collective
behaviour of active dipoles results in the appearance of mesoscopic
cooperative regions contributing to the macroscopic relaxation process
\cite{jonscher93,jonscher96}.
The number $L_{N}$ of such mesoscopic regions in the system is
determined by their sizes $M_{1}, M_{2},\ldots$ and by the number of
the active dipoles in the system \cite{kwaj2001}.

The effective relaxation rate $\tilde\beta_{N}$ 
consists of the contributions 
$\overline{\beta_{jN}}$ of
all $L_{N}$ mesoscopic cooperative regions
\begin{equation}
\label{21}
\tilde\beta_{N} = \sum\limits_{j=1}^{L_{N}} \overline{\beta_{jN}}. 
\end{equation}
Similarily, the relaxation
rate $\overline{\beta_{jN}}$ of the $j$th mesoscopic region 
is the sum of the contributions $\beta_{iN}$
of all active dipoles over the region [13, 32-34]
and hence it is equal to 
\begin{equation}
\label{22}
\overline{\beta_{jN}}=
\sum\limits_{i=M_{1}+\ldots+M_{j-1}+1}^{M_{1}+\ldots+M_{j}} \beta_{iN}.
\end{equation}
As mentioned in section 2 (i), the individual
relaxation rate, denoted here by $\beta_{iN}$, 
reflects the random intra-cluster
dynamics, i.e. statistical properties of the interactions of
the $i$th active dipole with some inactive neighbours 
forming around it a cluster of size $N_{i}$. 

Comparing (\ref{19})	and (\ref{21}) with $\overline{\beta_{jN}}$
of the form (\ref{22}) one obtains the explicit formula
for $\nu_{N}$, the number of the contributions to the total relaxation
rate, in the framework of the 
proposed approach. For the sake of this paper it is not
neccessary to cite the entire complicated formula derived (for details 
see \cite{kwaj2001}). It is only important to note that $\nu_{N}$ is fully 
determined by $N$, the system size; by the cluster sizes $N_{1},
N_{2},\ldots$; and by the cooperative-region sizes
$M_{1},M_{2},\ldots$.
The number of the dipoles directly engaged in the 
relaxation process, their location, interaction range, relaxation
rates, and all the quantities defined by them are random. 
Their precise stochastic characteristics determining the relaxation 
response of the system are unknown, unfortunately. However, 
as a rule, the relaxing systems consist of a large number of dipoles so
that the distribution of $\tilde\beta_{N}$ can be satisfactorily
approximated by the weak limit
\begin{equation}
\label{23}
\tilde\beta=\lim\limits_{N\to\infty} \tilde\beta_{N}.
\end{equation}
In practice, even $N\sim 10^{6}$ can suffice to replace
adequately $\tilde\beta_{N}$ in (\ref{20}) by the limit $\tilde\beta$.

On the basis of the limit theorems of probability theory the
distribution of the effective relaxation rate $\tilde\beta$ in
(\ref{23})
and the resulting form of the relaxation response (\ref{20}) in
time or the corresponding susceptibility in frequency domain
$$
\chi'(\omega)-i\chi''(\omega)\propto\phi^{*}(\omega)= 
 \int\limits_{0}^{\infty}e^{-i\omega
t}\left ( -\D{d\phi \over dt}(t)\right )\,dt
$$
can be derived even with rather 
restricted information on properties of the micro/mesoscopic
statistical levels in the system. It appears \cite{kwaj2001,ajkw2001} 
that each of the well-known dielectric responses can be obtained in this
way if appropriate conditions are imposed on the basic random
quantities: the cluster sizes $N_{i}$, the sizes $M_{j}$ of the
cooperative regions of correlated clusters, and the individual
active-dipole relaxation rates $\beta_{iN}$.

\section{Foundations of the universal response}
Table 2 collects the statistical properties shown
\cite{kwaj2001,ajkw2001} to underlie 
the KWW, HN, CC, CD, and Debye relaxation responses, given by formulas 
(\ref{KWW}) and (\ref{HN}). 
As it follows from the table, the empirical responses
are realized by various stochastic schemes. The conditions imposed on 
$N_{i}$, $M_{j}$, and $\beta_{iN}$ here take one of two forms; either the
distribution of the quantity considered is of finite mean value
or it has a heavy tail. Let us explain that 
the distribution of a nonnegative random variable, say $X$,
has a heavy tail if for some $0\!<\!a\!<\!1$
the tail function $\Pr (X\!>\!x)\sim x^{-a}$ for
large $x$, i.e., 
\begin{equation}
\label{ht}
\lim\limits_{x\to\infty}\D{\Pr (X\!>\!x)\over x^{-a}} = {\rm const} >0.
\end{equation}
Classical example of the heavy-tailed distribution is the Pareto
distribution \cite{johnson}
$$
\Pr (X\!>\!x)=\D{1\over 1+(Ax)^{c}},\;\;x>0
$$
with the shape parameter $0<c<1$ and any positive scale constant $A$. 
(In this case the heavy-tail exponent $a=c$.)
Another distribution possesing this property is a generalisation of the
Pareto example; namely, the Burr law \cite{johnson}
$$
\Pr (X\!>\!x)=\D{1\over (1+(Ax)^{c})^{d}},\;\;x>0
$$
with the shape parameters $c,d>0$ such that $0<cd<1$
and any positive scale constant~$A$. (Here the heavy-tail exponent
$a=cd$.)
More difficult but interesting heavy-tailed distributions 
are the completely asymmetric L{\'e}vy-stable laws \cite{feller}
defined by their Laplace transforms 
${\cal L}(X;t)=\exp(-(At)^{c})$	with $0<c<1$ and $A>0$.
(In this case the heavy-tail exponent $a=c$.)

Condition (\ref{ht}) applied to any random variable $X$ expresses
the following scaling property of the magnitude represented by $X$:
\begin{equation}
\label{25}
\Pr(X\geq Cx) \approx C^{-a}
\Pr(X\geq x), \;\; x\to \infty,
\end{equation}
for any fixed constant $C>0$.
(If there exists the corresponding probability density $w(x)$, the
condition (\ref{25}) yields $w(Cx)\sim C^{-a-1}w(x)$ for large $x$.)
Hence, in the presented approach the heavy-tail property
is directly related to the spatial 
(if referred to $N_{i}$ and $M_{j}$, the cluster and the cooperative-region 
sizes) or temporal (if referred to the relaxation rate
$\beta_{iN}$) scaling properties of the system. 
Comparing the non-Debye responses with the Debye one, as done in
Table~3, one can see that the difference between them lies in the
presence of such a scaling in the system. Namely, in the Debye case, 
that of the individual behaviour of the active dipoles, there is no property 
of this form.  By contrast, spatial or/and temporal scaling can be found for 
other responses. 
Moreover, at least one of the two variates $M_{j}$
and $\beta_{iN}$ has a heavy-tailed distribution in the non-Debye cases.
Since, as equation (\ref{22}) shows,
the correlated-cluster relaxation rate $\overline{\beta_{jN}}$ includes 
both those quantities; its
distribution has also a heavy tail, 
``producing'' a hierarchy of mesoscopic relaxation rates; i.e. 
the same proportion of smaller or larger mesoscopic contributions 
to the effective response no matter the scale at which one is 
looking at the relaxation rate distribution. This property follows 
from both the stochastic properties of the active dipoles generated by
the intra-cluster dynamics and the strength of screening yielding the 
range of interactions	between them.

As a consequence of the proposed approach, the origins of the
macroscopic energy criterion coefficient can be pointed out.
As we have already explained, 
the heavy-tailed distribution of any (micro/meso/macroscopic)
relaxation rate $\beta$ reflects the scaling property of the
rate
\begin{equation}
\label{26}
\Pr(\beta\geq Cb) \approx C^{-a}
\Pr(\beta\geq b), \;\; b\to \infty,
\end{equation}
for any fixed constant $C>0$ (where $a$ is the heavy-tail exponent). 
As in (\ref{10}) and (\ref{17}), we
have $\Pr (\theta\geq t) = \left \langle \exp (-\beta t)\right \rangle $
so that the properties of the micro/meso/macroscopic relaxation rates 
determine the survival
probabilities of the microscopic (active dipole), mesoscopic (cooperative 
region),
and macroscopic (entire system)	objects, respectively; thus, by analogy
to (\ref{20}), defining the function $\phi(t)=\left\langle
\exp(-\beta t) \right\rangle$, for simplicity referred to as
the micro/meso/macroscopic relaxation function.

The asymptotic behavior (\ref{26}) of the relaxation-rate distribution 
at large $b$ is connected with the short-time behavior of the associated
relaxation function $\phi(t)$ \cite{feller}. 
Namely, at the origin $t\to 0$ the response function
$f(t)=-\D{d\phi(t)\over dt}$ takes the form
\begin{equation}
\label{27}
f(t)\propto t^{a\!-\! 1} L(t)
\end{equation}
where $L(t)$ is a function slowly varying at $t=0$ (i.e.,
$L(Ct)/L(t)\to 1$ as $t\to 0$ for any fixed positive constant $C$). 
Then, it can be shown \cite{saichev} that the short--time property
(\ref{27}) of the response function $f(t)$ 
corresponds to the following asymptotic behavior of 
$\chi (\omega)\propto\phi^{*}(\omega)= 
 \int\limits_{0}^{\infty} e^{-i\omega t} f(t)dt$
for the high-frequency region:
\begin{equation}
\label{28}
\chi (\omega)=
\chi '(\omega)-i\chi ''(\omega) \propto
(i\omega)^{\T -a} L(1/\omega)\;\;\;{\rm for} \;\omega\to\infty.
\end{equation}
Property (\ref{28}) leads straightforwardly to the constant ratio
$$
\D{\chi  ''(\omega) \over \chi' (\omega)} = 
{\rm cot}\left (\eta\D{\pi\over 2}\right )
\;\;\;{\rm for}\;\omega\gg\omega _{ p}
$$
with $\eta=1-a$. This result, interpreted as the (micro/meso/macroscopic) 
energy criterion, is consistent with the energy criterion (\ref{encr}) if 
only $\eta$ is equal to the macroscopic high-frequency power-law exponent $n$. 
Summing up, the heavy-tail property of the
relaxation rate with the heavy-tail exponent $a$ leads to the 
energy criterion with the characteristic constant $\eta=1\!-\!a$. 

Table~4 contains the detailed discussion of the energy criterion at the
micro/meso/ macroscopic levels for the cases related to the KWW, HN, CC
and CD empirical relaxation responses (\ref{KWW}) and (\ref{HN}).
Observe that in each particular response the parameter $\eta$ followed from 
the mesoscopic-relaxation-rate tail is equal to the macroscopic 
energy-criterion coefficient $n$, and hence, that
the high-frequency universal laws (\ref{fpl}) and (\ref{encr}) and
its parameter $n$ are determined by the heavy-tail
properties of the mesoscopic (correlated-cluster) relaxation rate
distributions.
The above proves Jonscher's hypothesis introduced in his
screening theory of relaxation and stating that in order to observe
property (\ref{encr}) at the macroscopic level, the same property must
characterise the behaviour of the relaxing dipoles. As shown in Table~4,
in general, the energy-criterion parameter agrees
with $n$ from equation (\ref{encr}) on the mesoscopic level, only.

\section{Conclusions}
To our knowledge, this is the first attempt to relate rigorously the
widely observed universal macroscopic laws to the micro/mesoscopic 
statistical characteristics of the complex system, thereby establishing
a basis for the understanding of the stochastic origins of the relaxation 
phenomenon. The essential element which was required for the derivation
of these results is the strict probabilistic formalism in terms of which 
the dipolar screening and energy criterion concepts can be expressed. 

In terms of our analysis, the conditions for the
universal relaxation response may be stated as follows:

\begin{enumerate}
\renewcommand{\labelenumi}{(\roman{enumi})}

\item at the microscopic level, a random number of the active dipoles,
those that follow changes of the external field, is selected;
their individual relaxation rates are determined by the interactions
of the active dipoles 
with inactive neighbours forming random-sized clusters around them;

\item at the mesoscopic level, correlated-cluster regions of sizes 
depending on the strength of screening appear; the collective rate
of relaxation of the active dipoles in such a mesoscopic cooperative
region becomes correlated with the number of the active dipoles in
the region and with the stochastic properties of their individual
relaxation rates, see equation (\ref{22}); the mesoscopic 
correlated-cluster
relaxation rate combines both the spatial 
and the temporal scaling properties of the system;

\item at the macroscopic level, 
averaging over the number of effective contributions and their rates
leads to the universal relaxation for the entire system, giving law
(\ref{fpl});
the sufficient condition of this is that the high-frequency
energy criterion is satisfied at the mesoscopic level; the mesoscopic
energy criterion is provided by the heavy-tail property of the
correlated-cluster relaxation rates.
\end{enumerate}


\clearpage
\bigskip
\noindent
{\bf Table 1.}\\
\underline{Dipolar screening schemes.}\\

{\small 
\begin{table}[th]
\begin{center}
{\large\bf Table 1}\\[2ex]
\begin{tabular}{||c|c|c|c|c||}
\hline\hline
\parbox{20mm}{\begin{center}
{\bf Condition}
\end{center}}&{\bf Screening}&{\bf Physical picture}&
\multicolumn{2}{c||}{\bf Consequences}\\
\hline\hline
$N_{d}\gg N_{1}$& effective & \parbox{35mm}{\begin{center}
Dipoles do not\\ ''see'' 
one another\end{center}}&\parbox{25mm}{\begin{center}
Individual\\ behavior\end{center}}&
\parbox{25mm}{\begin{center}Debye-like\\response\end{center}}\\
\hline
$N_{d}\ll N_{1}$& ineffective & \parbox{35mm}{\begin{center}Dipoles interact\\
strongly\end{center}}&
\parbox{25mm}{\begin{center}Collective\\ behavior\end{center}}&
\parbox{25mm}{\begin{center}Universal\\response\end{center}}\\
\hline\hline
\end{tabular}
\end{center}
\end{table}
}

\clearpage
\bigskip
\noindent
{\bf Table 2.}\\
\underline{Stochastic origins of the empirical relaxation responses.}\\

{\small
\begin{table}[th]
\begin{center}
{\large\bf Table 2}\\[2ex]
\begin{tabular}{||c|c|c||c|c||}
\hline\hline
\multicolumn{3}{||c||}{\raisebox{-3ex}[-1ex][-1ex]{\bf Conditions for:}}&
\raisebox{-3ex}[-1ex][-1ex]{\parbox{22mm}{\begin{center}
{\bf Relaxation\\response}
\end{center}}}&
{\raisebox{-3ex}[-1ex][-1ex]
{\parbox{22mm}{\begin{center}{\bf Energy-criterion\\parameter}
\end{center}}}}\\
\multicolumn{3}{||c||}{}&&\\
\multicolumn{3}{||c||}{}&&\\
\cline{1-3}
\cline{5-5}
\parbox{22mm}{\begin{center}
active-dipole\\relaxation\\rate $\beta_{iN}$
\end{center}}&
\parbox{22mm}{\begin{center}
cluster\\size $N_{i}$
\end{center}}&
\parbox{22mm}{\begin{center}
cooperative\\-region\\size $M_{j}$
\end{center}}&&
\parbox{17mm}{\begin{center}
$0\!<\!n\!<\! 1$
\end{center}}\\
\hline\hline
\parbox{22mm}{\begin{center}
heavy tail\\with $a\!=\!\alpha$
\end{center}}&
\parbox{22mm}{\begin{center}
heavy tail\\with $a\!=\!\alpha$
\end{center}}&
\parbox{22mm}{\begin{center}
heavy tail\\with $a\!=\!\gamma$
\end{center}}&
\parbox{22mm}{\begin{center}
{\bf HN}\\$0\!<\!\alpha\!<\!1$\\$0\!<\!\gamma\!<\!1$
\end{center}}&
\parbox{17mm}{\begin{center}
$n\!=\!1\!-\!\alpha\gamma$
\end{center}}\\
\hline
\parbox{22mm}{\begin{center}
heavy tail\\with $a\!=\!\alpha$
\end{center}}&
\parbox{22mm}{\begin{center}
heavy tail\\with $a\!=\!\alpha$
\end{center}}&
\parbox{22mm}{\begin{center}
$\langle M_{j}\rangle\!<\!\infty$
\end{center}}&
\parbox{22mm}{\begin{center}
{\bf CC}\\$0\!<\!\alpha\!<\!1$\\$\gamma\!=\!1$
\end{center}}&
\parbox{17mm}{\begin{center}
$n\!=\!1-\alpha$
\end{center}}\\
\hline
\parbox{22mm}{\begin{center}
$\langle \beta_{iN}\rangle\!<\!\infty$
\end{center}}&
\parbox{22mm}{\begin{center}
$\langle N_{i}\rangle\!<\!\infty$
\end{center}}&
\parbox{22mm}{\begin{center}
heavy tail\\with $a\!=\!\gamma$
\end{center}}&
\parbox{22mm}{\begin{center}
{\bf CD}\\$\alpha\!=\!1$\\$0\!<\!\gamma\!<\!1$
\end{center}}&
\parbox{17mm}{\begin{center}
$n\!=\!1\!-\!\gamma$
\end{center}}\\
\hline
\parbox{22mm}{\begin{center}
heavy tail\\with $a\!=\!\alpha$
\end{center}}&
\parbox{22mm}{\begin{center}
$\langle N_{i}\rangle\!<\!\infty$
\end{center}}&
\parbox{22mm}{\begin{center}
$\langle M_{j}\rangle\!<\!\infty$
\end{center}}&
\parbox{22mm}{\begin{center}
{\bf KWW}\\$0\!<\!\alpha\!<\!1$
\end{center}}&
\parbox{17mm}{\begin{center}
$n\!=\!1-\alpha$
\end{center}}\\
\hline
\parbox{22mm}{\begin{center}
$\langle \beta_{iN}\rangle\!<\!\infty$
\end{center}}&
\parbox{22mm}{\begin{center}
$\langle N_{i}\rangle\!<\!\infty$
\end{center}}&
\parbox{22mm}{\begin{center}
$\langle M_{j}\rangle\!<\!\infty$
\end{center}}&
\parbox{22mm}{\begin{center}
{\bf Debye}\\$\alpha\!=\!1$\\$\gamma\!=\!1$
\end{center}}&
\parbox{17mm}{\begin{center}
energy\\criterion\\not fulfilled
\end{center}}\\
\hline\hline
\end{tabular}
\end{center}
\end{table}
}

\clearpage
\bigskip
\noindent
{\bf Table 3.}\\
\underline{Comparison of the non-Debye relaxation responses to the Debye
one.}\\

{\small
\begin{table}[th]
\begin{center}
{\large\bf Table 3}\\[1ex]
\begin{tabular}{||c|c||}
\hline\hline
&\\
\begin{minipage}[t]{75mm}
\begin{center}
{\bf Non-Debye responses}\\
(collective behavior of the dipoles):
\end{center}
\end{minipage}&
\begin{minipage}[t]{75mm}
\begin{center}
{\bf Debye response }\\
(individual behavior of the dipoles):
\end{center}
\end{minipage}\\
&\\
\hline\hline
&\\
\begin{minipage}[t]{75mm}
\begin{center}
The relaxation rates of the dipoles responding to
changes of the external electric field,
\end{center}
\end{minipage}
&
\begin{minipage}[t]{75mm}
\begin{center}
The relaxation rates of the dipoles responding to
changes of the external electric field,
\end{center}
\end{minipage}\\
&\\
\begin{minipage}[t]{75mm}
\begin{center}
{\bf or}
\end{center}
\begin{center}
the cluster sizes,
\end{center} 
\end{minipage}
&
\begin{minipage}[t]{75mm}
\begin{center}
{\bf and}
\end{center}
\begin{center}
the cluster sizes,
\end{center}
\end{minipage}\\
&\\
\begin{minipage}[t]{75mm}
\begin{center}
{\bf or}
\end{center}
\begin{center}
the sizes of correlated-cluster regions
\end{center}
\begin{center}
{\small\bf have heavy-tailed
distributions,\\ and, as a result,\\ infinite expected values.}
\end{center}
\end{minipage}
&
\begin{minipage}[t]{75mm}
\begin{center}
{\bf and}
\end{center}
\begin{center}
the sizes of correlated-cluster regions\\
(if any)
\end{center}
\begin{center}
{\small\bf have distributions\\ with finite mean values.}
\end{center}
\end{minipage}\\
&\\
\begin{minipage}[t]{75mm}
\rule{20mm}{0.25pt}\\
{\scriptsize (Note that the HN response is the only case in which all three
distributions have heavy tails.)}
\end{minipage}
&
\begin{minipage}[t]{75mm}
\rule{20mm}{0.25pt}\\
{\scriptsize (Note that {\scriptsize
\it the dipoles do not have to respond with
the same rate} but the rates are of limited distribution.)}
\end{minipage}\\
&\\
\hline
&\\
\begin{minipage}[t]{75mm}
Spatial and/or temporal scaling is present 
at the different statistical levels of the system.
\end{minipage}
&
\begin{minipage}[t]{75mm}
Neither spatial nor temporal scaling appears at any level.
\end{minipage}\\
&\\
\hline\hline
\end{tabular}
\end{center}
\end{table}
}

\clearpage
\bigskip
\noindent
{\bf Table 4.}\\
\underline{Discussion of the energy criterion at the micro/meso/macroscopic
levels.}

{\small 
\begin{table}[th]
\begin{center}
{\large\bf Table 4}\\[2ex]
\begin{tabular}{||c||c|c|c||c|c|c||}
\hline\hline
\raisebox{-5ex}[-1ex][-1ex]{\parbox{18mm}{\begin{center}
{\bf Universal\\response}
\end{center}}}& 
\multicolumn{3}{c||}{
\raisebox{-3ex}[-1ex][-1ex]{\bf Relaxation rate}}
&\multicolumn{3}{c||}{\raisebox{-3ex}[-1ex][-1ex]{
\boldmath$\chi ''(\omega) / \chi ' (\omega) = 
{\bf cot}\left (\eta\pi/ 2\right )$}\unboldmath}\\
&\multicolumn{3}{|c||}{}&\multicolumn{3}{|c||}{}\\
&\multicolumn{3}{|c||}{}&\multicolumn{3}{|c||}{}\\
\cline{2-7}
&\parbox{20mm}{\begin{center}
$\beta_{iN}$\\(micro)
\end{center}}&
\parbox{20mm}{\begin{center}
$\overline{\beta_{jN}}$\\(meso)
\end{center}}&
\parbox{20mm}{\begin{center}
$\tilde\beta$\\(macro)
\end{center}}&
\parbox{23mm}{\begin{center}micro\end{center}}
&\parbox{20mm}{\begin{center}meso\end{center}}
&\parbox{20mm}{\begin{center}macro\end{center}}\\
\hline\hline
{\bf HN}&\parbox{20mm}{\begin{center}
heavy tail\\with $a\!=\!\alpha$
\end{center}}&
\parbox{20mm}{\begin{center}
heavy tail\\with $a\!=\!\alpha\gamma$
\end{center}}&
\parbox{20mm}{\begin{center}
heavy tail\\with $a\!=\!\alpha\gamma$
\end{center}}&
$\eta=1\!-\!\alpha\neq n$
&\multicolumn{2}{c||}{
$\eta=1\!-\!\alpha\gamma=n$
}\\
\hline			   	
\parbox{18mm}{\begin{center}
{\bf CC}\\{\bf KWW}\end{center}}
&\parbox{20mm}{\begin{center}
heavy tail\\with $a\!=\!\alpha$
\end{center}}&
\parbox{20mm}{\begin{center}
heavy tail\\with $a\!=\!\alpha$
\end{center}}&
\parbox{20mm}{\begin{center}
heavy tail\\with $a\!=\!\alpha$
\end{center}}&\multicolumn{3}{c||}{
$\eta=1\!-\!\alpha=n$
}\\
\hline
{\bf CD}&\parbox{20mm}{\begin{center}
$\langle \beta_{iN}\rangle\!<\!\infty$
\end{center}}&
\parbox{20mm}{\begin{center}
heavy tail\\with $a\!=\!\gamma$
\end{center}}&
\parbox{20mm}{\begin{center}
heavy tail\\with $a\!=\!\gamma$
\end{center}}&
\parbox{23mm}{\begin{center}
does not\\hold
\end{center}}
&\multicolumn{2}{c||}{
$\eta=1\!-\!\gamma=n$
}\\
\hline\hline
\end{tabular}
\end{center}
\end{table}
}


\begin{figure}[p]
\centerline{\epsfxsize=8cm \epsfbox{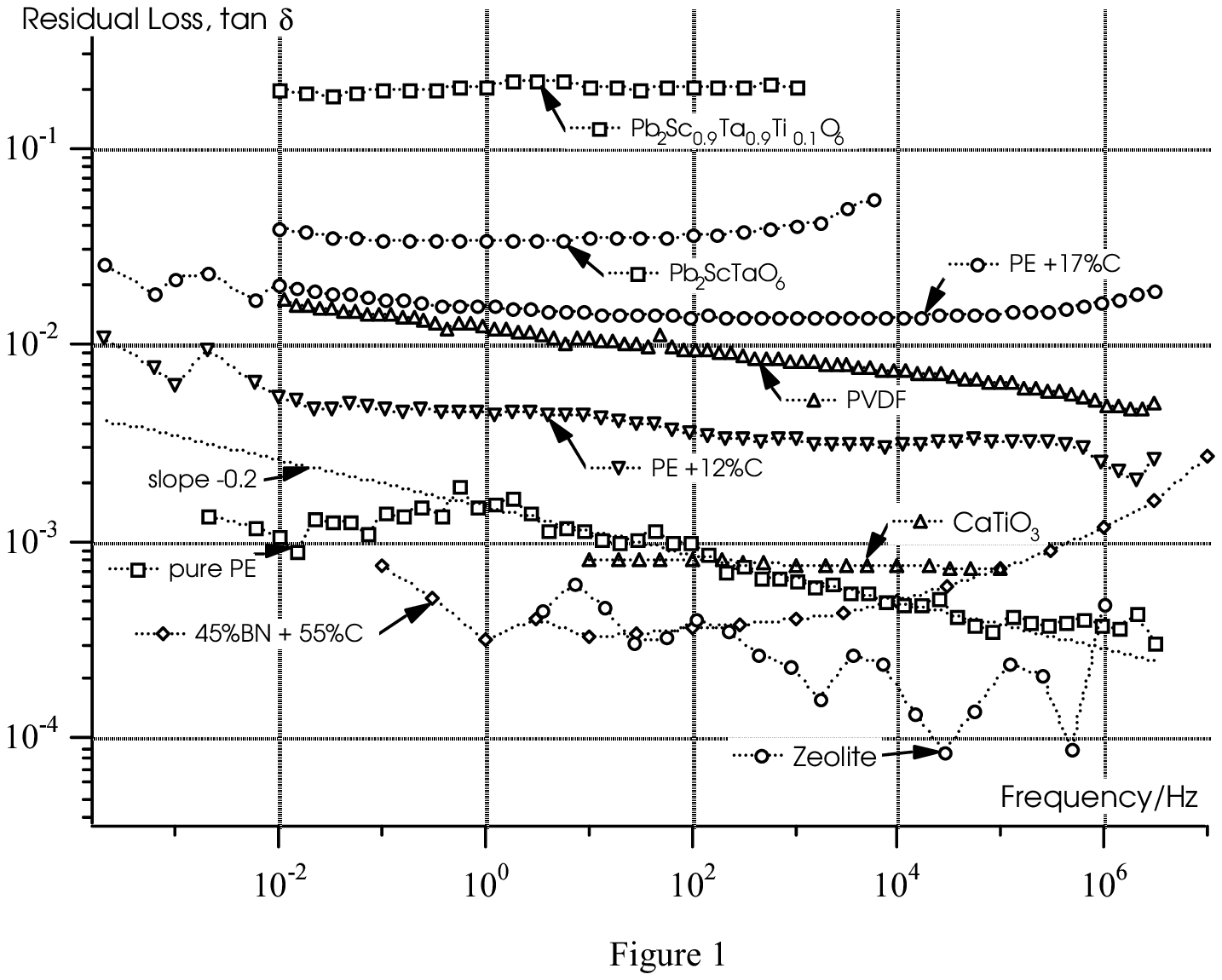} }
\end{figure}

\newpage
\noindent
{\large\bf Figure captions} \\[5pt]
\bigskip
\noindent
{\bf Figure 1.}\\
Typical examples of the frequency dependence of the residual loss, after
substraction of dc component, low-frequency dispersion and strong
dipolar peaks, plotted logarithmically against frequency, for a range
of dielectric materials. The data cover more than ten decades of
frequency, and the prevailing trend is a ``flat'' or
frequency-independent loss with only minor deviations, for instance the
slope $-0.2$ over six decades for pure polyethylene. The $\tan\delta$
values cover a range of barely three decades between the lowest and
highest losses; and the surprising feature is the very slight effect of
the addition of conducting species (like 12\% graphite to polyethylene
raising the loss by a factor of 3, and 17\% of graphite by a further
factor of 3) with very little change of the frequency dependence.
Likewise, the addition of 55\% of graphite to boron nitride does not 
raise the
extremely low loss of the system. Most of the data quoted here are
taken from much wider ranges for variable temperature and for ranges of
composition. The evident trends are the general ``flatness'' of the
losses  in frequency which is not compatible with any known
mechanism and that irrespective  of the absolute level of loss. 
Detailed information: $\rm Pb_{2}Sc_{0.9}Ta_{0.9}Ti_{0.1}O_{6}$ 
and $\rm Pb_{2}ScTaO_{6}$ are ferroelectric ceramics, 
from A.~Isnin and A.K.~Jonscher, Ferroelectrics, 
210 (1998) 47. Pure polyethylene (PE) showing nine decades of frequency
with a very slowly varying $\tan\delta$	in the range $3\times 10^{-4}$
- $2\times 10^{-3}$ with a frequency dependence proportional to
$f^{-0.2}$ over most of the middle range; also shown are data for
polyethylene with admixtures of 12\% and 17\%  of graphite;
from D.S.~McLachlan and M.B.~Heaney, Phys.~Rev.~B 60
(1999) 12746. Single crystal $\rm CaTiO_{3}$ over a range of
temperatures with $\tan\delta = 5 -9\times 10^{-4}$, show extremely
small values of $1\! -\! n= 0.000325$ to $0.000625$; 
from B.S.~Lim, A.V.~Vaysleyb, and A.S.Nowick, Applied Physics
A 56 (1993) 8. Alpha-poly-vinylidene di-fluoride (PVDF$\alpha$) data 
from A.K.~Jonscher and G.~Menegotto, 
IEEE Trans.~Dielectrics and EI 7 (2000) 303.
Percolation system graphite-boron nitride (45\%BN+55\%C) from 
J.J.~Wu and D.S.~McLachlan, Phys.~Rev.~B 56 (1997) 1236;
Phys.~Rev.~B 58 (1998) 14880.
Zeolite plus NLO-active
para-aniline from M.~Wuebbenhorst, private communication.

\newpage
\noindent
{\bf Prof.~Andrew K.~Jonscher:}	Born 1922 Warsaw. 1944 Joined the
Polish Army in Italy. Engineering studies in London 1945; B.Sc.~(Eng)
1st class Hons 1948; Ph.D.~in semiconductors from Queen Mary College,
University of London, 1951. 1952-1963 on the Staff of the Hirst
Research Center of the General Electric Company in Wembley, England,
finishing as Leading Scientific Staff. 1963-1965 Reader in Solid State
Physics at Chelsea College, University of London; 1965-1985 Professor
of Solid State Electronics there. 1970 established Chelsea Dielectrics
Group, one of the leading groups working on dielectric relaxation in
solids, formulated the ``universal'' dielectric response, known in the
literature as Jonscher's law. 1983 set up Chelsea Dielectrics Press to
publish his first monograph ``Dielectric Relaxation in Solids'' which
was too ``hot'' for commercial publishers to touch. 1987 reached
retirement age and transferred as Visiting Professor to Royal Holloway,
University of London, where he continued experimental and theoretical
work on relaxation. 1990 nominated Whitehead Memorial Lecturer of IEEE
Electrical Insulation Society. 1996 Chelsea Dielectrics Press publishes
his second monograph ``Universal Relaxation Law''. 1998 Honorary
Doctorate of the Technical University in {\L}{\'o}d{\'z}, Poland, for ``pioneering
work on dielectric relaxation''. Most recently theoretical studies of
relaxation jointly with the Wroc{\l}aw group.\\[1cm]
{\bf Dr Agnieszka Jurlewicz:} 
Ph.D.~in Mathematics  (1994), Institute of Mathematics, 
Faculty of Basic Problems of Technology, 
Wroc{\l}aw University of~Technology.
Since October 1995 Assistant Professor at the Institute of Mathematics,
Wroc{\l}aw University of Technology. \underline{Research
topics:} 
Theoretical investigations of a new type of continuous-time random
walk; application of the obtained results in the probabilistic model of
relaxation phenomena in disordered systems; studies on the relationship 
between the model and the Adam-Gibbs theory.\\[1cm] 
{\bf Prof.~Karina Weron:} 
Ph.D.~in Theoretical Condensed Matter Physics (1978), 
Faculty of Basic Problems of Technology, 
Wroc{\l}aw University of~Technology.
Since May 2001 Full Professor at the Institute of Physics,
Wroc{\l}aw University of Technology. \underline{Research
topics:} Statistical and stochastic methods in physics (random
matrices, random walks, self-similar stochastic processes, time series
analysis); modelling of relaxation phenomena; dynamical properties of
complex systems; ionic transport through biological membranes.
\end{document}